\documentclass[conference]{IEEEtran}
\usepackage{color}
\usepackage{epsfig}
\usepackage{amsmath,amssymb}
\usepackage{setspace}
\usepackage{slashbox}
\usepackage{tikz}
\usetikzlibrary{shapes,arrows}
\usepackage{caption}
\usepackage{subcaption}
\usepackage{threeparttable}
\usepackage{multirow}

\begin{document}
\title{Evaluation of 5G New Radio Non-orthogonal Multiple Access Methods for Military Applications}

\author{\IEEEauthorblockN{Cagri Goken}
\IEEEauthorblockA{ASELSAN Inc., Ankara, Turkey\\
Email: cgoken@aselsan.com.tr}
\and
\IEEEauthorblockN{Alptekin Yilmaz}
\IEEEauthorblockA{ASELSAN Inc., Ankara, Turkey\\
	Email: alptekiny@aselsan.com.tr}
\and
\IEEEauthorblockN{Onur Dizdar}
\IEEEauthorblockA{ASELSAN Inc., Ankara, Turkey\\
	Email: odizdar@aselsan.com.tr}
}
\maketitle

\begin{abstract}
Non-orthogonal multiple access (NOMA) is an enabling technique to support massive connectivity and utilize the radio resources more efficiently. A number of novel NOMA schemes have been proposed for 5G New Radio (NR) standards. In this study, we evaluate various 5G NOMA methods for different military communications scenarios. First, we provide the description of basic principles in each evaluated scheme, then we investigate and compare their performances under different system parameters such as spectral efficiency, overload factor and antenna numbers in various channel models. Finally, we provide the discussions and insights on the suitability of the evaluated schemes for the considered military scenarios based on simulations. 	
	
\end{abstract}
\begin{IEEEkeywords}
	Non-orthogonal multiple access (NOMA), 5G, satellite communications, military networks, polar codes.
\end{IEEEkeywords}

\IEEEpeerreviewmaketitle

\section{Introduction and Motivation} \label{sec:Intro}

NOMA schemes are based on the idea that multiple users share the same resource block (e.g. time slot, subcarrier group) via non-orthogonal resource allocation. The main motivation behind NOMA is to increase system capacity by utilizing the resources more efficiently and/or provide enhanced connectivity \cite{SPfor5G}. The idea of users sharing the same resource blocks is not new and has been used in previous commercial wireless systems, e.g. 3G, and military waveforms, Mobile User Objective System (MUOS) for military UHF satellite communication (SATCOM) \cite{MUOS}. Starting with the standardization process of 5G waveform, there has been a renewed interest in NOMA recently. 3GPP has evaluated a number of novel NOMA schemes extensively targeting  massive machine type communications (mMTC). This use case requires the connection of massive number of low-cost, energy-efficient devices sending sparse, small packets in the uplink communications \cite{3GPP_Final}. There has not been a final conclusion about which techique(s) will be used in future releases (Rel. 16 and onward) despite promising results yet.

Most of the recently proposed NOMA methods are considered for commercial scenarios, NOMA can also be employed in modern military communications systems as well. Some of the example scenarios where NOMA methods can provide significant benefits in military communications can be listed as:

\begin{figure}[t!]
	\begin{subfigure}[t]{0.48\textwidth}
		\centering
		\hspace{0.3cm}\includegraphics[scale=0.27]{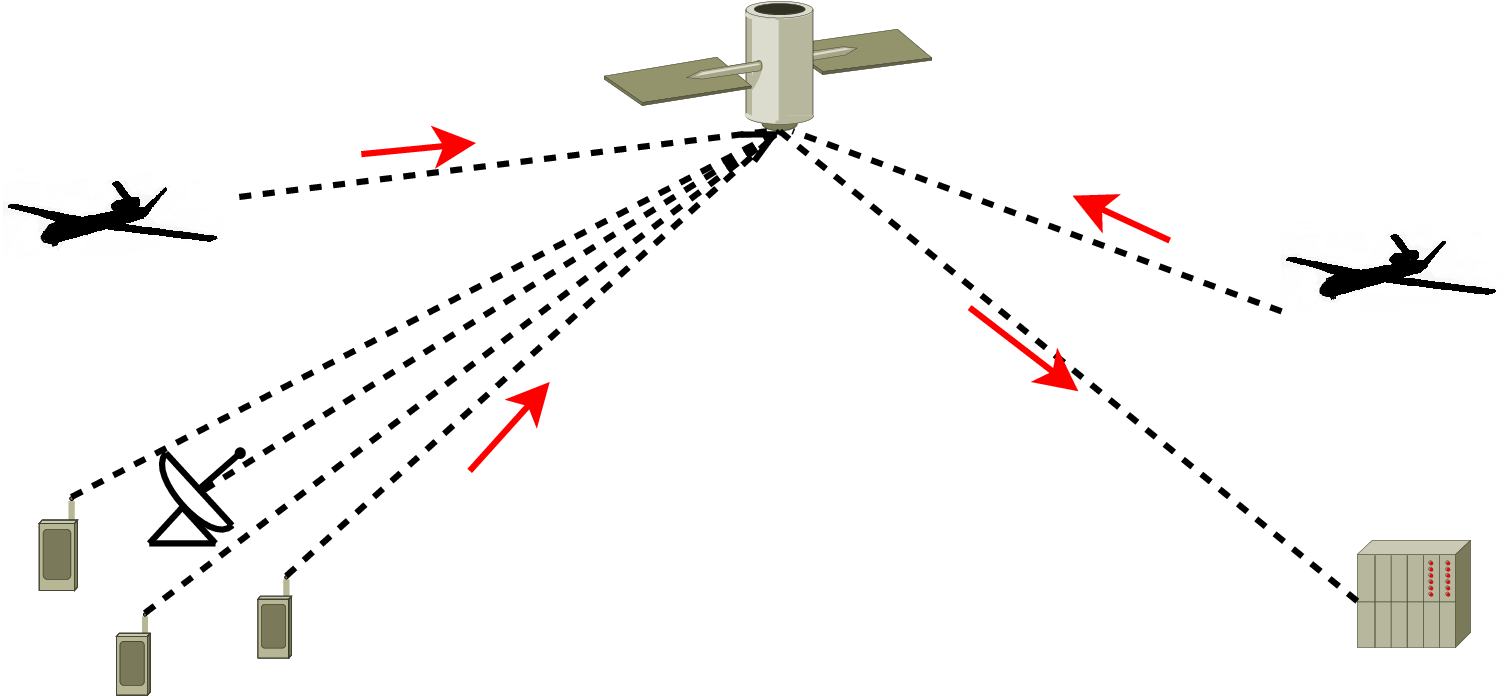}
		\caption{Uplink satellite communications} \label{fig:sat}
	\end{subfigure}
	\hfill
	\vspace{-0.5cm}
	\centering
	\begin{subfigure}[b!]{0.22\textwidth}
		\centering
		\vspace{0.7cm}
		\hspace{-0.6cm}\includegraphics[scale=0.15]{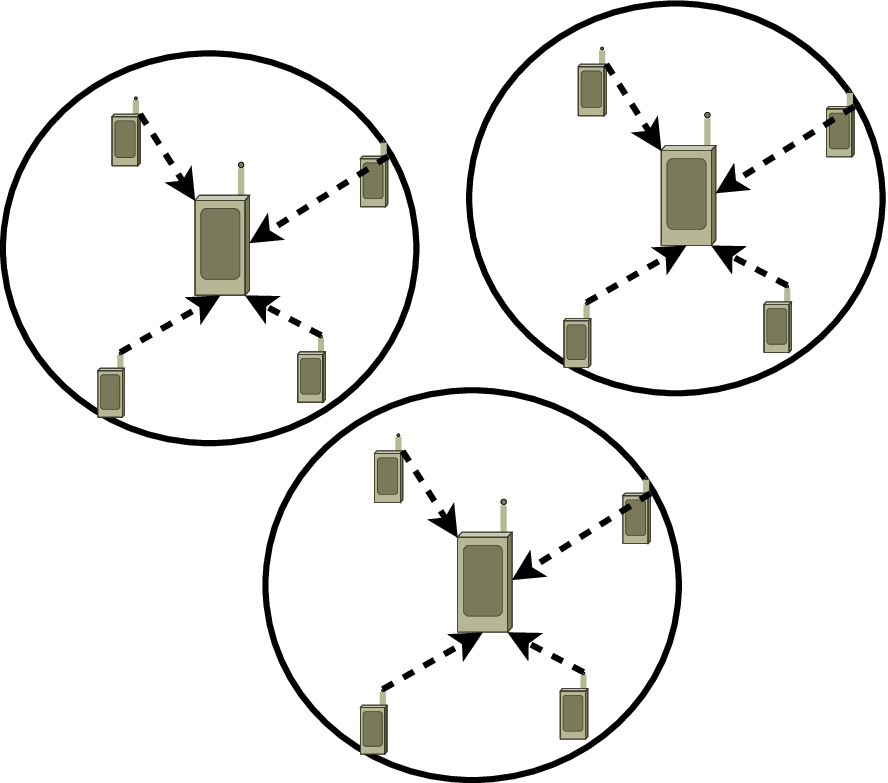}
		\caption{Cluster/contention-based \\ communications} \label{fig:cluster}
	\end{subfigure}
	\begin{subfigure}[b!]{0.2\textwidth}
		\centering
		\vspace{0.7cm}
		\includegraphics[scale=0.24]{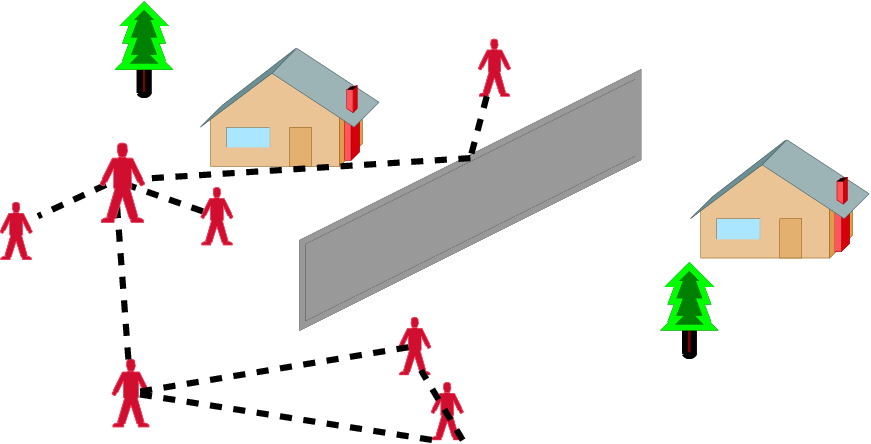}
		\vspace{0.1cm}
		\caption{Close-range/indoor \\ communications} \label{fig:indoor}
	\end{subfigure}
	\caption{Military scenarios potentially employing NOMA}
	\vspace{-0.7cm}
	\label{fig:new_f_functions}
\end{figure}
\vspace{-0.3cm}
\begin{itemize}
		\item In satellite communications, the efficient use of bandwidth becomes more and more necessary with the increasing number of terminals and unmanned aerial vehicles (UAVs) using the satellite links. NOMA can be used for channel requests and synchronization signaling as an alternative to dynamical channel allocation methods by reducing delay and overhead.
		\item In covert military operations, the soldiers can send various types of data supplied by sensors and cameras to the team leader in close-range and indoor communications scenarios \cite{Covert}.  NOMA techniques can be employed in such scenarios to send information from large number of monitoring devices to a fusion/command center.
		\item In tactical area communications, the increased number of connected radios in a network lead to a clustering based network structure rather than a flat network \cite{clustering_survey}. For the uplink communications inside the cluster, members can send packets to the cluster-head by NOMA schemes,  which will potentially increase the total spectral efficiency and simplify the link scheduling algorithms. Also, NOMA can be used instead of contention based access methods used for channel allocation requests to reduce packet collisions and increase the system capacity.
	\end{itemize}
    
Even though a wide variety of NOMA techniques are evaluated in the literature, the main evaluation scenario of the proposed techniques is communications with low spectral efficiency in channels with slow fading and short RMS delay spread ($30-300$ ns.) as it targets commercial platforms. In military communications, the channel conditions and use cases can be quite different from the commercial ones. The robustness and reliability of the communications even in rough channel conditions are the top priorities. Therefore, a new technique has to be tested thoroughly and its characteristics and limits need to be well-understood. Therefore, it is important to analyse the different NOMA schemes under diverse scenarios to decide about their suitability to aforementioned possible military use cases. Our main contribution in this study is to test and evaluate the performances of NOMA methods in different channel models and use cases in the scope of military communications, which has not been considered in the literature to the best of our knowledge. Our goal is to understand the characteristics of each method under varying conditions and draw useful conclusions based on extensive simulations. In order to fulfill this purpose, we have selected a candidate technique from different NOMA classes. We have also analyzed the performance of superposed Orthogonal Frequency Division Multiplexing (OFDM) as the benchmark scheme, in which no user-specific signatures are employed and the user signals are separated by advanced successive-interference cancelation (SIC) receiver. We provide our main observations in the discussion part in Section IV.

\begin{figure*}
	\centering
	\vspace{-1cm}
	\hbox{\hspace{2.5em} \includegraphics[scale=0.3]{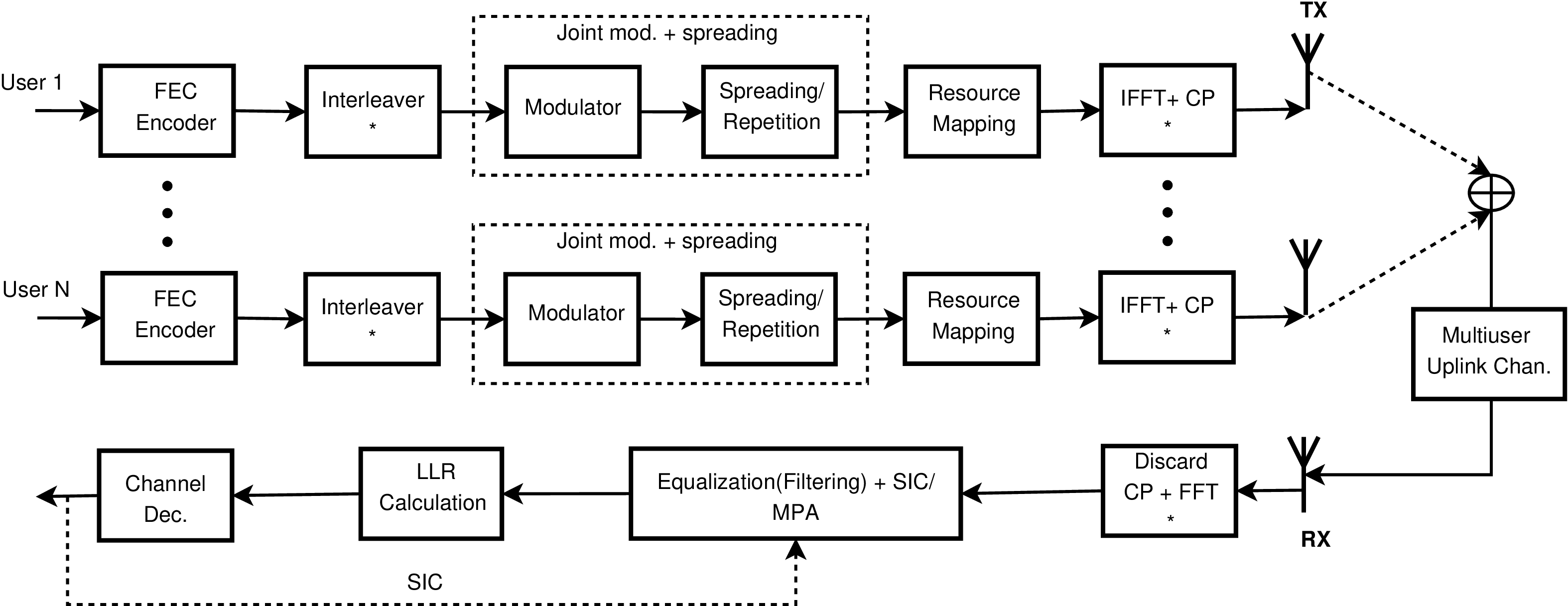}}
	\caption{NOMA TX-RX.} \label{fig:NOMATXRX}
\end{figure*}

\section{Background}\label{sec:Background}

\subsection{System Model} \label{sec:System}

In this paper, we consider uplink transmission scenarios with OFDM, in which $M$ users share the same physical (time and frequency) resources to communicate with a single receiver. The transmissions of the users are assumed to be synchronized at the receiver. 
The received signal at the $k$-th subcarrier, $k \in \left\lbrace 0, 1, \ldots, N-1 \right\rbrace $, is modeled as
\begin{align} \label{eq:model}
Y_{k}=\sum_{i=1}^{M}\sqrt{p_{i}}h_{i,k}s_{i,k}+n_{k}, 
\end{align}
where $n_{k}$ is the additive white Gaussian noise at the receiver, $p_{i}$ is the transmitted power for the $i$-th user, $h_{i,k}$ is the combined effect of path loss and fading and $s_{i,k}$ is the transmitted symbol (including effects of NOMA processing) at the $k$-th subcarrier for the $i$-th user, respectively. 

\subsection{Basic Principles} \label{sec:Basics}

The novel NOMA schemes proposed for 5G share the common idea to superpose different users' signals in the same orthogonal resource block in a controlled manner so that they can be recovered using advanced receiver structures. The ratio of the number of users to the number of resource blocks is called the \textit{overload factor}. In order to limit the multi-user interference and distinguish between the users, user specific signatures need to be used. The proposed NOMA schemes for 5G are classified depending on the type of signatures, which can be in power-domain or in modulation and symbol level processing including spreading, repetition, interleaving and codebook mapping \cite{3gpp_NOMA3}. 

A general description of NOMA transmitter and receiver\footnote{The blocks with * are optional in some schemes, even though they are included in this study.} structures considered in this study are depicted in Fig. \ref{fig:NOMATXRX}. At the transmitter side, the encoded user data are modulated and repeated/spread with user-specific sequences. Then, the user signals are superposed in specific resource blocks. 
 It should be noted that modulation and spreading operations are performed jointly in some of the considered NOMA schemes. 

At the receiver side, advanced receiver structures, such as Successive Interference Cancellation (SIC) or Message Passing Algorithm (MPA) are employed depending on NOMA schemes to recover user data under multi-user interference. In SIC, the received signal is first filtered by Matched Filter (MF) or  Minimum Mean Squared Error (MMSE) filter. For soft input FEC decoding, the log-likelihood ratios (LLR) are calculated in accordance with \cite{llr_calculation}. 
After LLR calculation, the user data are first decoded, then reconstructed and subtracted from the received signal until all user data has been processed. The processing order of users can be formed according to their post filter signal to noise-interference ratio (SINR) or any other criteria. This procedure can be performed for several iterations if necessary. On the other hand, MPA is a generic algorithm providing near Maximum Likelihood (ML) detection performance and working on a factor graph representation. It calculates the probability of the codewords for each user by iteratively passing messages in a bipartite graph.  

\subsection{Evaluated Methods} \label{sec:Methods}

In this study, we evaluate 4 different 5G NOMA schemes in addition to superposed OFDM as the benchmark scheme. We briefly describe each method as follows:

\subsubsection{Sparse Coded Multiple Access (SCMA)} \label{sec:SCMA}
SCMA is a codebook-based NOMA method which performs modulation and spreading jointly \cite{scma1}. Particularly, each user has a different codebook which contains $J$ different sparse codewords with length $N$. At the transmitter, a user maps $\log_2 J$ bits to a specific codeword in the codebook and codewords are sparse in the sense that the number of nonzero elements in a codeword is much less than $N$. 
The sparse structure of SCMA codewords allows MPA to be a suitable receiver to detect user data. For SCMA, the codebook available in \cite{scma2} has been used in this study. 

\subsubsection{Pattern Division Multiple Access (PDMA)} \label{sec: PDMA}
PDMA is a codebook-based NOMA scheme employing a user-specific pattern mapping of the modulated user symbols to resource blocks for differentiating the user data. PDMA patterns are selected to offer different orders of transmit diversity. To recover the user data, a receiver based on MMSE filter working together with SIC algorithm iteratively or MPA is proposed in the literature \cite{PDMA_TransComm}, \cite{PDMA_Conf}.

\subsubsection{Repetition Division Multiple Access (RDMA)} \label{sec: RDMA}
RDMA is an interleaver based NOMA scheme introduced in \cite{RDMA3GPP}. The scheme separates the user data and utilizes time and frequency diversity by assigning distinct cyclic shift repetition patterns in frequency domain to users. Due to the RDMA pattern, the interference level between users can be limited and randomized so that an SIC receiver with MF or MMSE filtering can distinguish the user data.

\subsubsection{Multi User Shared Access (MUSA)} \label{sec: MUSA}

MUSA is a spreading sequence-based NOMA method as proposed in \cite{MUSAConf}. In this method, each user's modulated symbols are spread by short, user-specific codes and transmitted in the same resource blocks. 
At the receiver, each user's data can be detected using SIC procedure. In \cite{MUSA3GPP}, a receiver structure employing CRC-based SIC and user-specific MMSE filtering is proposed and it has been used in this study as well. Also, we use the spreading sequences given in \cite{MUSA3GPP}.

\subsubsection{Superposed OFDM (S-OFDM)} \label{sec: PCBMA}

{S-OFDM can be considered as a simple NOMA scheme where the user signals utiliizing OFDM signaling are simply added to each other in the same resource blocks without any NOMA type processing at the transmitter side and SIC is utilized at the receiver side. 
A very detailed complexity analysis has been provided in \cite{3GPP_Final}. Therefore, we mainly focus on the error performances of the studied schemes due to lack of space, which are provided in the next section.

\section{Numerical Results}\label{sec:NumRes}

\subsection{Scenarios}  \label{sec:Scenarios}

In satellite communications, high Doppler and phase errors can be observed due to the orbital movement of satellites.  Additionally, creating significant received power difference in is difficult since variations in the user channels are limited \cite{corazza_1994}, \cite{patzold_1998} and performing back-off at significant levels is not feasible due to large communication distances. Accordingly, we assume that the received power differences between the multiplexing users are chosen independently from a uniform distribution, i.e. \eqref{eq:model} becomes $Y_{k}= \sum_{i=1}^{M} \alpha_i s_{i,k}+n_{k}, $ where \mbox{$10\log\left(|\alpha_i|^2 \right)\sim U(-l,+l)$}. 

For sensor networks/military IoT and covert operation/indoor communications, the environmental conditions are similar to those in mMTC scenario. For example, in \cite{Covert}, statistics for RMS delay spread for soldier to soldier links are provided where most of the energy of the signal arrives in $25$ ns, and occasional multipath components arrive between $100$ and $150$ ns. Therefore, for close-range/indoor communications and sensor networks, we use TDL-A channel which has an RMS delay spread of $30$ ns to evaluate NOMA schemes, which is also one of the selected channels for the evaluation of NOMA schemes in 5G \cite{3gpp_NOMA1}.

For tactical area communications, various channel models can be adopted depending on the considered scenario. In this work, we assume a communications requirement in a densely built urban area and consider the COST207 Bad Urban (BU) wideband channel model \cite{wiley_book}. 

\begin{table}[hbt!]
\footnotesize	
\caption{Simulation Parameters} 
\centering 
\begin{threeparttable}
\begin{tabular}{c |c |c }
\hline\hline 
 FFT Size  & \multicolumn{2}{c}{$512$}       \\  \hline
\# of OFDM Subcarriers  & \multicolumn{2}{c}{$256$}    \\  \hline 
OFDM Symbol Length [us] &  \multicolumn{2}{c}{$60$}     \\   \hline
OFDM CP Length [us] &  \multicolumn{2}{c}{$15$}      \\   \hline
Carrier Frequency [MHz] & \multicolumn{2}{c}{$2000$}																\\   \hline
Modulation  & \multicolumn{2}{c}{QPSK}																\\   \hline
FEC  & \multicolumn{2}{c}{Polar}																\\   \hline
OFDM Symbol Number       & \multicolumn{2}{c}{$4$} 												\\   \hline
													& RDMA  &	\multirow{4}{*}{$512$}										\\   
													 & SCMA &																\\   
Codeword Length [bits]		 & PDMA &																	\\   
													& MUSA &																	\\   \cline{2-3}
													 & S-OFDM &    $2048$															\\		\hline
Spectral Efficiency					& \multicolumn{2}{c}{$1/4$, $1/6$}						\\    \hline
CRC Length [bits] &  \multicolumn{2}{c}{$16$}   												\\  \hline 
FEC Decoder  & \multicolumn{2}{c}{SCL-$16$}																\\   \hline
Antenna Number  & \multicolumn{2}{c}{$1$$\times$$1$ (SISO), $1$$\times$$2$ (SIMO)} 												\\   \hline
Channel Estimation &  \multicolumn{2}{c}{Ideal, MMSE}  								\\   \hline
\multirow{6}{*}{Receiver}  & RDMA &	MF-SIC-CRC										\\   \cline{2-3}
													 & SCMA &	MPA														\\   \cline{2-3}
													 & PDMA &	MMSE-SIC-CRC										\\   \cline{2-3}
													 & MUSA &	MMSE-SIC-CRC															\\   \cline{2-3}
													 & S-OFDM & MMSE-SIC-CRC														\\		\hline
Monte Carlo &  \multicolumn{2}{c}{$10000$}   												\\  \hline 													 
\end{tabular}
\end{threeparttable}
\label{table:sim_parameters} 
\end{table} 
    
\subsection{Simulation Results}\label{sec:SimRes}
The system parameters and the receiver structures for the schemes are given in Table~\ref{table:sim_parameters}. First, we test the NOMA schemes described in Section \ref{sec:Methods} with $150\%$ and $300\%$ overload factors. \footnote{Note that SCMA can support a maximum number of 6 users using 4 resources, therefore we evaluated SCMA in only $150\%$ overload factor. One can notice that the codeword length for S-OFDM is $4$ times those of other schemes. S-OFDM does not perform any repetition or spreading on multiple physical resources as other NOMA schemes so that we employ a longer codeword for S-OFDM over $4$ OFDM symbols for the sake of fairness.} In the simulations, the user spectral efficiencies (SE) are set to $1/4$ and $1/6$ bits/s/Hz for all evaluated schemes. The channel estimation is assumed to be ideal in all cases but the scenario with CFO for which MMSE based channel estimation is performed. 

In Figures~\ref{fig:AWGN_Sce_1} and \ref{fig:AWGN_Sce_2}, the performances of the considered NOMA schemes with $150 \%$ overload factor are provided for satellite communications scenario with $l=1$ and $1/4$ and $1/6$ bits/s/Hz SE per user, respectively. It is observed that the schemes RDMA and S-OFDM do not perform well, when the SE per user is $1/4$. On the other hand, all schemes except S-OFDM perform well with varying successes when the SE per user is decreased down to $1/6$. The results show that uplink NOMA methods are required in scenarios without significant channel variations as S-OFDM performs poorly in all of them. Also, the performances of the schemes are improved when SE is decreased. In Fig. \ref{fig:AWGN_Sce_3}, we observe the effects of CFO errors when the SE per user is $1/6$. We note the significant performance loss as compared to the case without CFO. Furthermore, RDMA does not work in this particular scenario.  These imply that CFO estimation and correction are necessary in general.

\begin{figure}[t!]
	\vspace{-0.5cm}
	\centering
	\hbox{\hspace{-1.3cm} \includegraphics[scale=0.41]{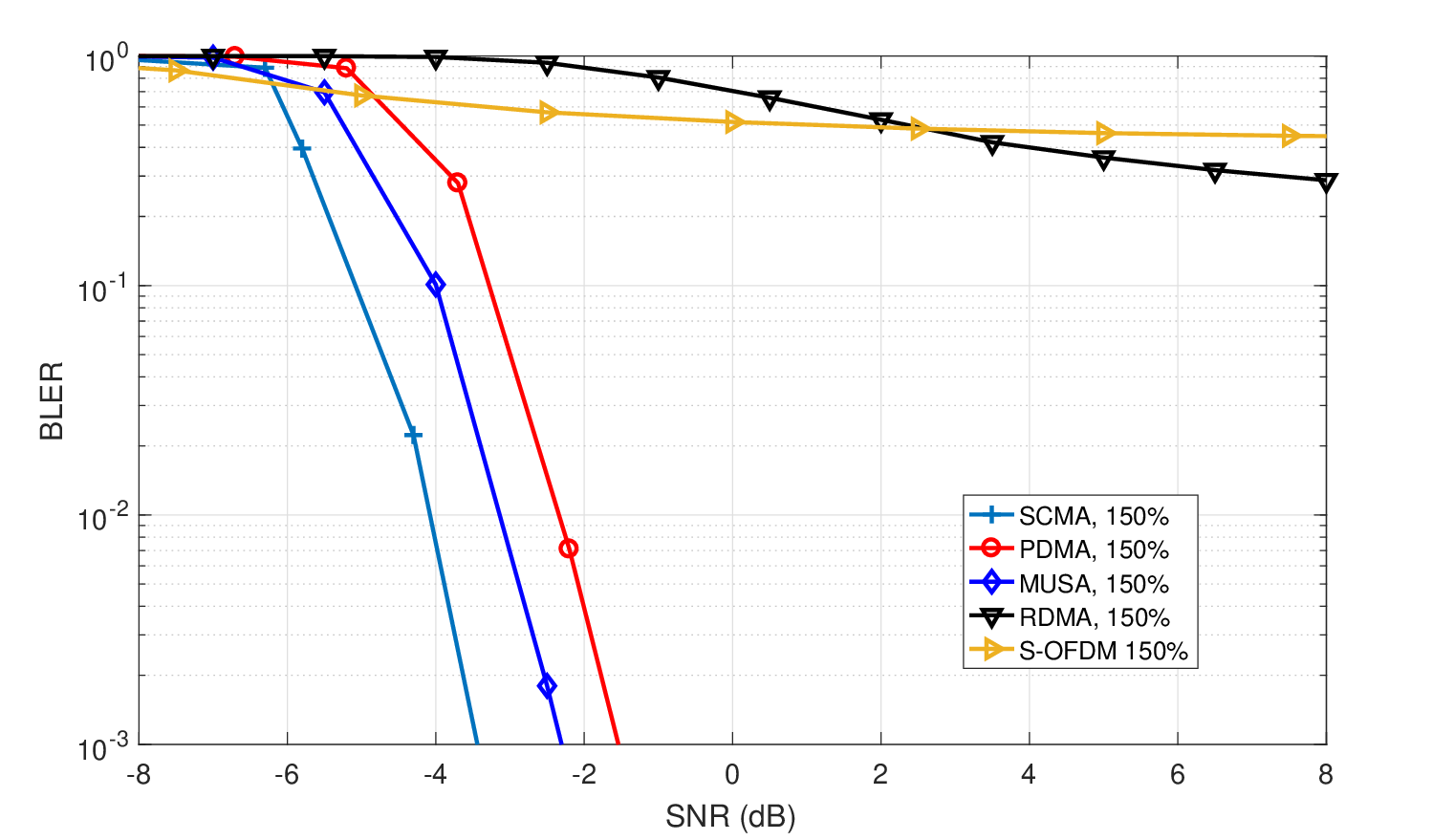}}
	\caption{\mbox{Scenario 1: $10\log\left(|\alpha_i|^2\right)\sim U(-1,+1)$}, SE per user = $1/4$ bits/s/Hz, SIMO.} 
	\vspace{-0.2cm}
	\label{fig:AWGN_Sce_1}
\end{figure}

\begin{figure}[t!]
	\vspace{-0.35cm}
	\centering
	\hbox{\hspace{-1.3cm} \includegraphics[scale=0.41]{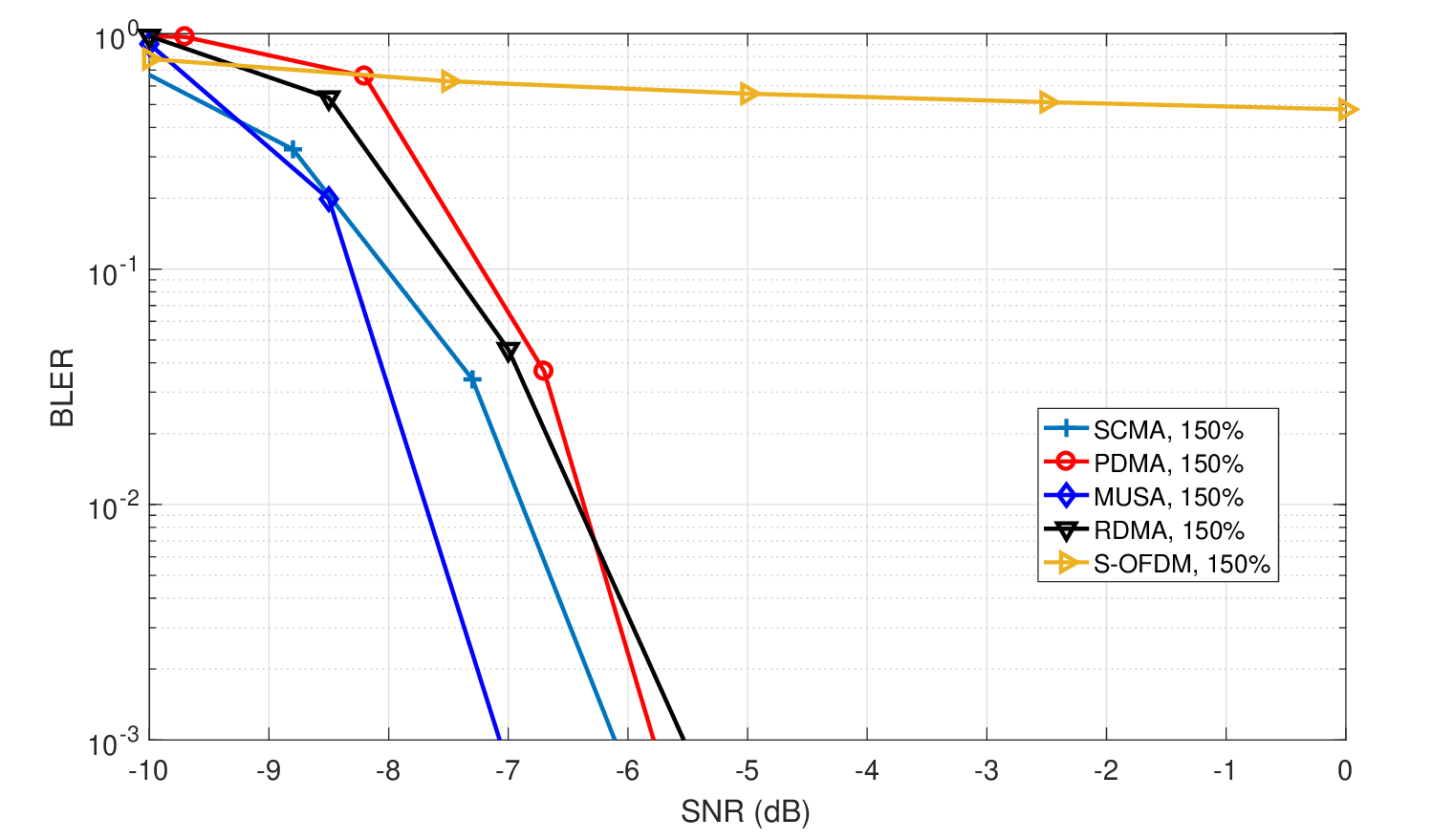}}
	\caption{\mbox{Scenario 2: $10\log\left(|\alpha_i|^2\right)\sim U(-1,+1)$}, SE per user= $1/6$ bits/s/Hz, SIMO.} 
	\label{fig:AWGN_Sce_2}
	\vspace{-0.6cm}
\end{figure}

In Figures~\ref{fig:BU_Sce_1} and \ref{fig:BU_Sce_2}, we investigate the performances of the schemes in BU channel for $1/4$ and $1/6$ bits/s/Hz SE per user, respectively. In this case, we assume that $p_{i}$ in \eqref{eq:model} are the same for all users in order to investigate the effect of channel selectivity on the error performance. We observe that the schemes perform much better compared to the case with small channel variations given above. The performance improvement is due to the fact that BU is a highly frequency selective channel, and it differentiates the multiplexing users via frequency diversity. We note that in both figures, S-OFDM gives the best performance as it is able to utilize lower code rates for a given SE per user compared to other schemes due to lack of operations such as repetition or spreading, hence it has a significant coding gain in this particular scenario.

\begin{figure}[t!]
	\vspace{-0.5cm}
	\centering
	\hbox{\hspace{-0.4cm} \includegraphics[scale=0.41]{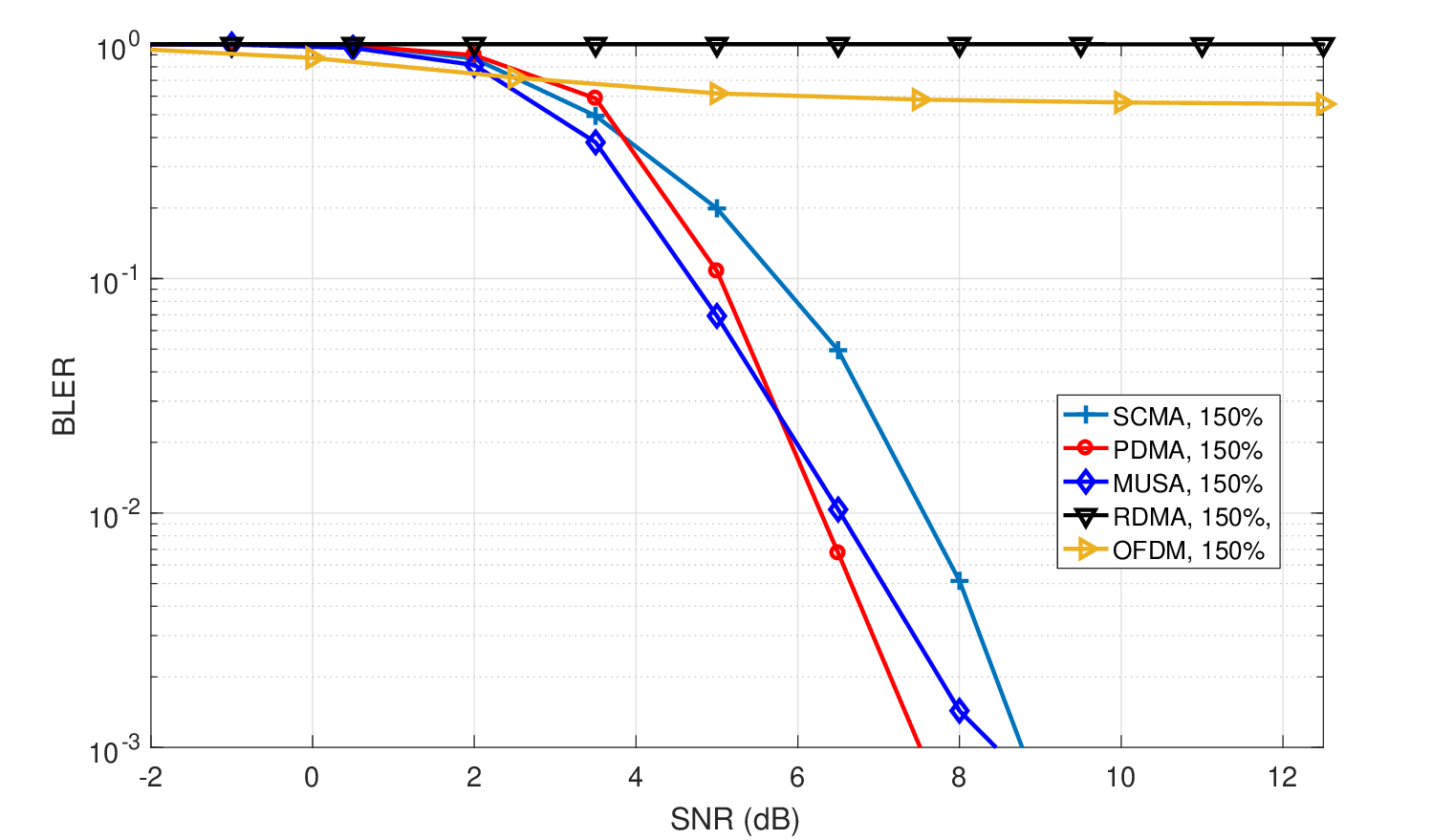}}
	\caption{\mbox{Scenario 3: $10\log\left(|\alpha_i|^2\right)\sim U(-1,+1)$}, CFO error uniformly distributed in $[-0.1,~+0.1]$ ppm, SE per user= $1/6$ bits/s/Hz, SIMO.}
	\label{fig:AWGN_Sce_3}
	\vspace{-0.2cm}
\end{figure}
	
\begin{figure}[t!]
	\vspace{-0.25cm}
	\centering
	\hbox{\hspace{-0.5cm} \includegraphics[scale=0.41]{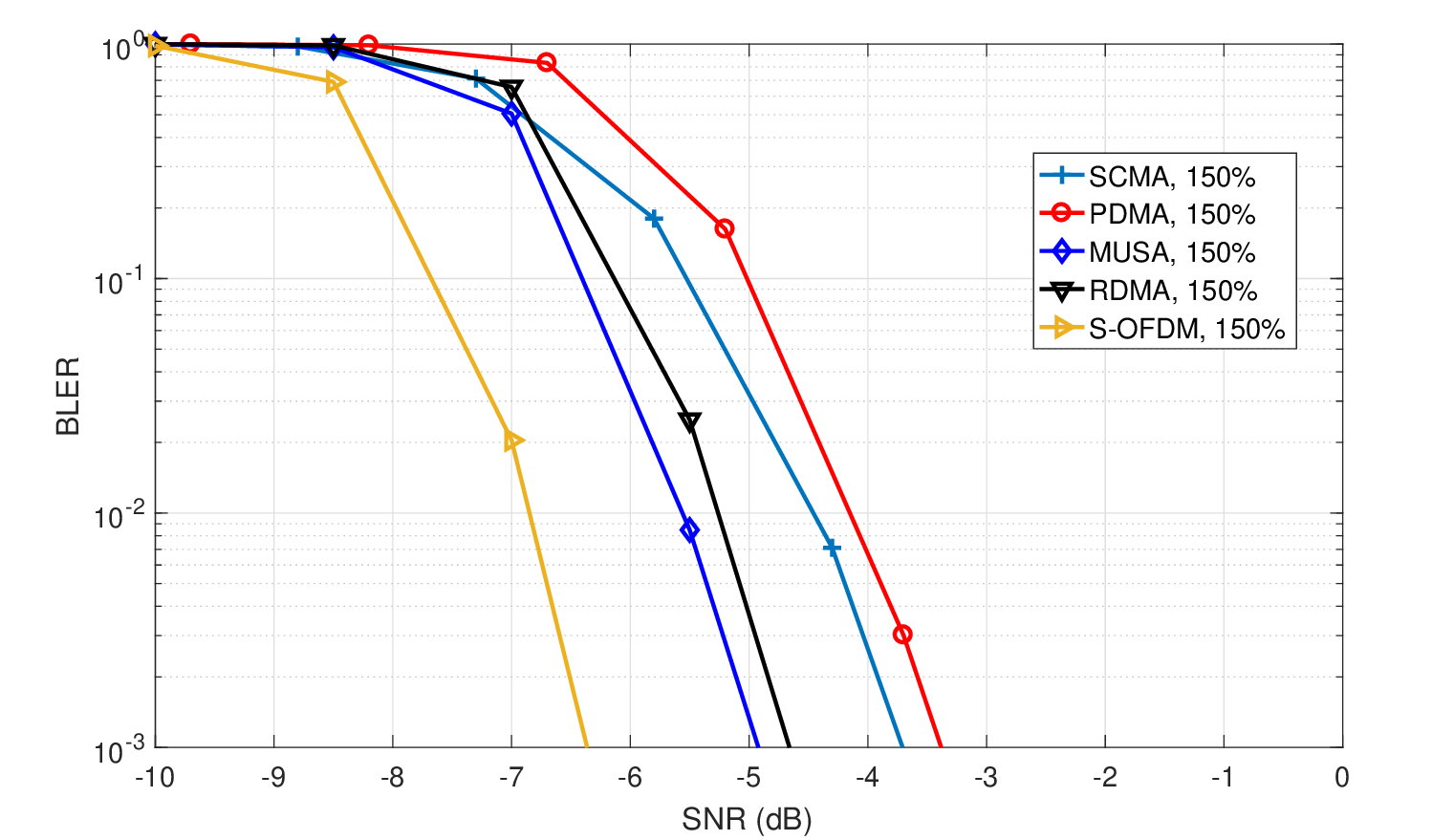}}
	\caption{Scenario 4: BU channel, SE per user= $1/4$ bits/s/Hz, SIMO.} 
	\label{fig:BU_Sce_1}
	\vspace{-0.4cm}
\end{figure}		

In Fig.~\ref{fig:TDLA_SIMO}, we investigate the performances of the schemes in TDL-A channel with $1/4$ bits/s/Hz SE per user. Again, we assume that $p_{i}$ are the same for all users. Note that TDL-A is a less frequency selective channel than BU. The diversity difference in the channels can be observed from the slopes of BLER curves in Figures~\ref{fig:BU_Sce_1} and \ref{fig:TDLA_SIMO}. However, TDL-A channel naturally provides a received power difference, which enables easier separation of users compared to the previously investigated cases, hence the performance of best and worst scheme is in a $3$~dB range. 

So far, we only considered SIMO configuration in our simulations. In Fig.~\ref{fig:TDLA_SISO}, we focus on a SISO setting which may be the case for certain military communications scenarios. The diversity gain by multiple receive antennas can be observed by comparing Figures~\ref{fig:TDLA_SIMO} and \ref{fig:TDLA_SISO}. The results in Fig.~\ref{fig:TDLA_SISO} imply that extra receiver antennas help the system, however it is still possible to apply NOMA schemes in a SISO system.

In Table~\ref{table:ResultsMethod}, we provide the SNR values to reach BLER target of $10^{-2}$ for $150\%$ and $300\%$ overload factors and all scenarios considered in the figures. In the Scenario 1, it is observed that none of the schemes can operate in high overload conditions. On the other hand, when the SE per user is decreased in the Scenario 2, MUSA and PDMA can reach the target BLER levels for $300 \%$ overload factor with 10-15 dB loss compared to $150 \%$ overload factor. We also note that in the Scenario 3, none of the schemes can reach the target level due to CFO present in the signal.

\begin{figure}[t!]
	\centering
	\vspace{-0.5cm}
	\hbox{\hspace{-1.2cm} \includegraphics[scale=0.41]{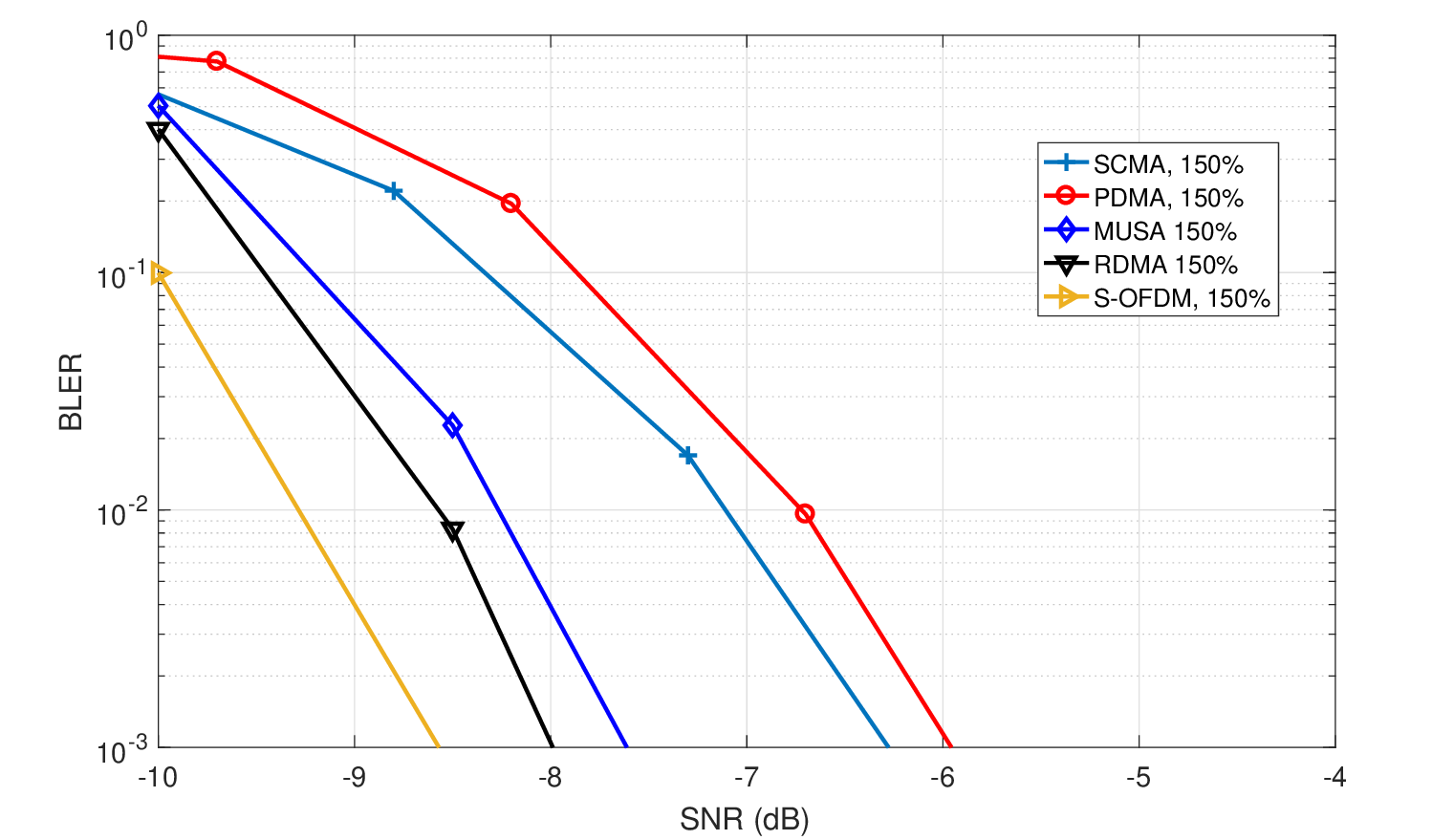}}
	\caption{Scenario 5: BU channel, SE per user= $1/6$ bits/s/Hz, SIMO.} 
	\label{fig:BU_Sce_2}
\end{figure}

\begin{figure} [t!]
	\vspace{-0.6cm}
	\centering
	\hbox{\hspace{-1.2cm} \includegraphics[scale=0.4]{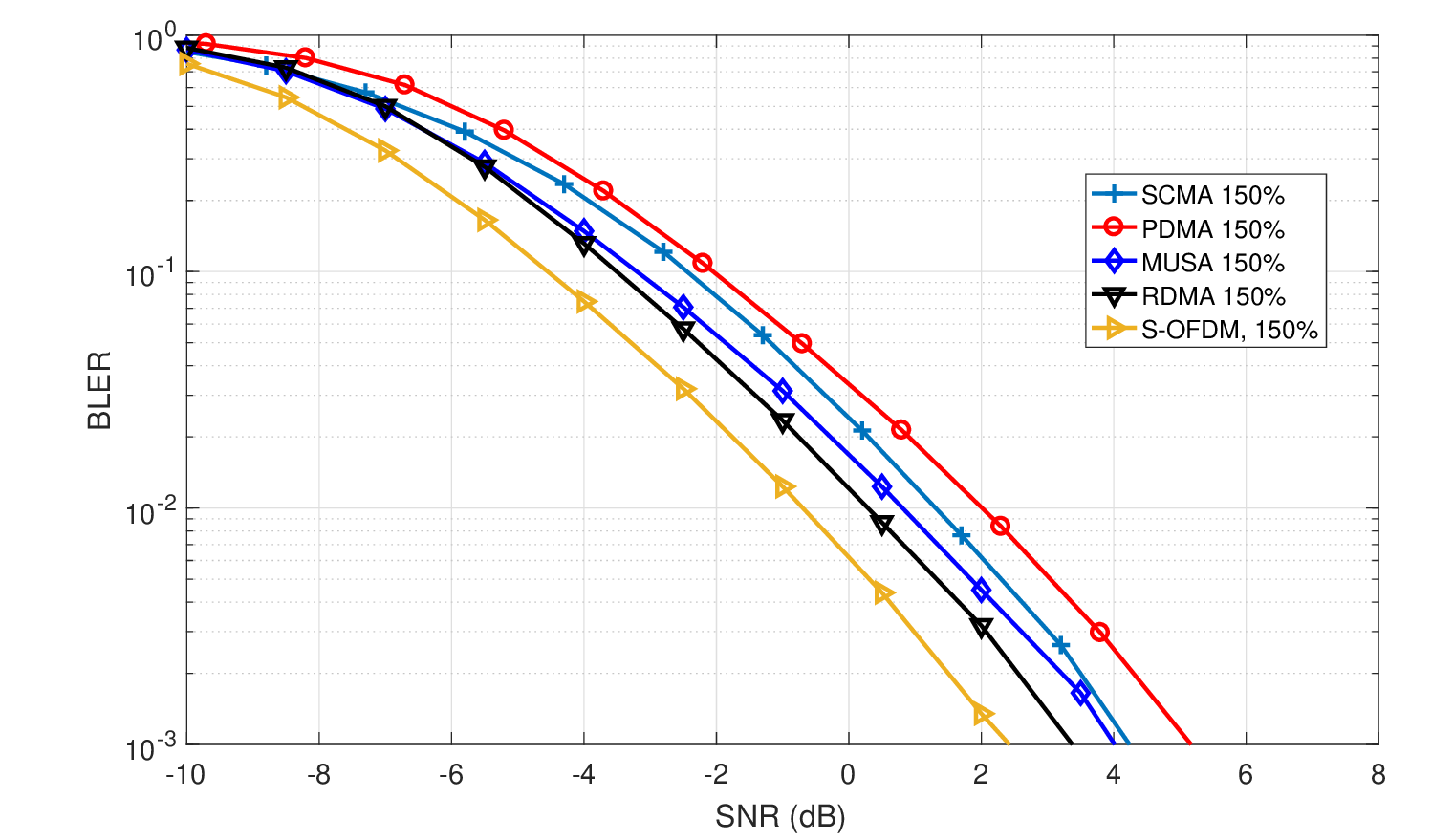}}
	\caption{Scenario 6: TDL-A channel, SE per user= $1/4$ bits/s/Hz, SIMO.} 
	\label{fig:TDLA_SIMO}
\end{figure}	

\begin{figure}[t!]
	\vspace{-0.6cm}
	\centering
	\hbox{\hspace{-1.2cm} \includegraphics[scale=0.41]{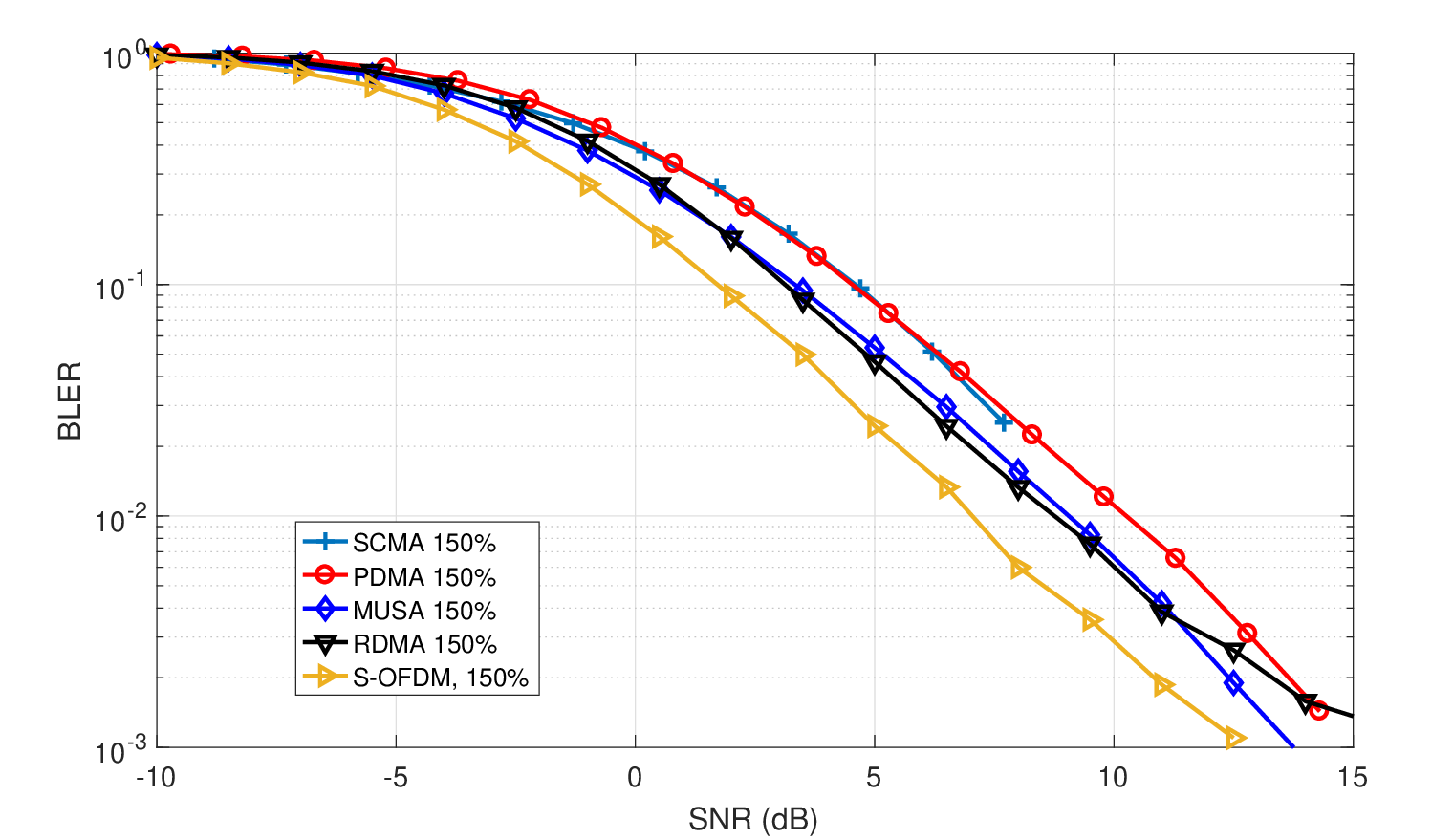}}
	\caption{Scenario 7: TDL-A channel, SE per user= $1/4$ bits/s/Hz, SISO.} 
	\label{fig:TDLA_SISO}
	\vspace{-0.5cm}
\end{figure}

In the Scenario 4, the performances of the MUSA and PDMA can surpass that of S-OFDM when overload factor is increased to $300\%$. This implies that these methods are more robust against overloading compared to S-OFDM in this scenario (see also, Fig.~\ref{fig:Overload_Combined}). However, when SE per user is decreased in the Scenario 5,  S-OFDM can still perform better than other schemes at both overload factors,. Also note that all schemes can reach the target BLER in this case.  

In the Scenario 6, high overload factors can be supported in TDL-A channel. For example, we observe that there is negligible performance gap ($\approx 0.05$ dB) between $150\%$ and $300\%$ overload factors for MUSA and S-OFDM. An important observation for the SISO case in the Scenario 7, is the performance degradation with $300\%$ overload in all schemes showing that using more receiver antennas is desirable from overloading perspective.

\begin{table}[t!] 
	\scriptsize
	\caption{SNR performances of the methods at target BLER [dB]} 
	\centering 
	\begin{tabular}{ c | c |c |c |c |c |c |c |c}
		\hline 
		\multirow{2}{*}{Scn.} &\multicolumn{2}{c}{MUSA}\vline &  \multicolumn{2}{c}{PDMA} \vline& \multicolumn{2}{c}{RDMA}\vline & \multicolumn{2}{c} {S-OFDM} 
		\\  \cline{2-9}
		& 150\% & 300\% & 150\% & 300\% & 150\% & 300\% & 150\% & 300\%
		\\ 
		\hline
		1 & -3.13  & N/A & -2.34 & N/A & N/A & N/A & N/A & N/A \\ \hline
		2 & -7.7  & 4.3 & -6.37 & 8.23 & -6.42 & N/A& N/A & N/A \\
		\hline
		3 & 6.52 & N/A & 7.48 &N/A& N/A & N/A & N/A & N/A \\
		\hline
		4 & -5.56 & -3.45 & -4.15 & -1.31 & -5.26 & N/A & -6.85 & -0.2 \\
		\hline
		5 & -8.26 & -7.92 & -6.72 & -3.9 & -8.57 & -5.58 & -9.27 & -8.14 \\
		\hline
		6 & 0.80 & 0.85 & 2.00 & 4.06 & 0.28 & N/A & -0.7 & -0.64 \\
		\hline
		7 & 9.1 & N/A & 10.25 & N/A & 8.75 &N/A & 7.05 & N/A\\
		\hline		
		
	\end{tabular}
	
	\label{table:ResultsMethod} 
\end{table} 

Next, we investigate the BLER performances of the schemes with respect to overload factors when SE per user= $1/4$ bits/s/Hz. In Fig.~\ref{fig:Overload_Combined}, we give the performance results for $3$ different channel scenarios. Note that, in Figures~\ref{fig:AWGN_Sce_1}--\ref{fig:AWGN_Sce_3}, we investigated the case with \mbox{$10\log\left(|\alpha_{i}|^{2}\right)\sim U(-1,+1)$} and it was observed that high overload factors can not be supported in this specific case. In Fig.~\ref{fig:Overload_Combined}, we consider a scenario with \mbox{$10\log\left(|\alpha_i|^{2}\right)\sim U(-7,+17)$}. One can observe that, MUSA outperforms S-OFDM and RDMA for all considered overload factors. Also, the performances of the schemes for TDL-A and BU channels are provided at SNR values of $2$~dB and $-2$dB, respectively. Note that in both channels, MUSA gives the best performance in high overload values. One can observe that the performance of MUSA is almost identical up to $500 \%$ overload, showing its robustness to high interference in TDL-A channel. In BU channel, it can be observed that the performances of all schemes degrade as the overload factor increases. Note that it is possible to support larger overload factors in TDL-A channel compared to BU. One can note that even though the performance of S-OFDM is the better than those of other schemes with low overload factors, its performance is quickly surpassed as the overload factor increases. The reason behind that is the fact that with low overload factors, using low code rates instead of repetition/spreading based methods provides coding gain; however as the overload factor increases, specific NOMA processing techniques help decrease the effects of interference. Therefore lowering the code rate in high overload scenarios is not a solution anymore, and NOMA methods are required.

\begin{figure}[t!]
	\vspace{-0.9cm}
	\centering
	\hbox{\hspace{-0.5cm} \includegraphics[scale=0.65]{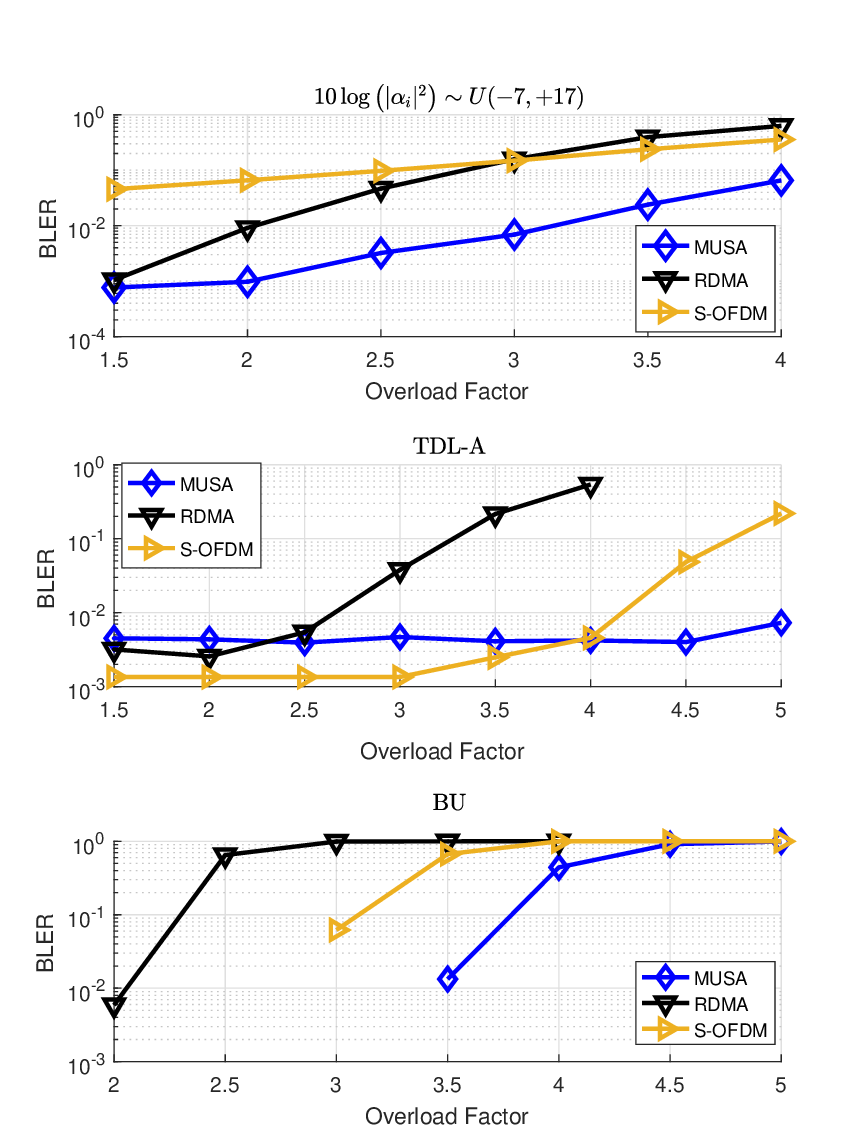}}
	\caption{BLER vs Overload Factor, SE per user= $1/4$ bits/s/Hz, SIMO.} 
	\label{fig:Overload_Combined}
	\vspace{-0.5cm}
\end{figure}

 



\section{Discussion and Conclusion}\label{sec:Conc}

In this study, we investigated the performances of 5G NOMA schemes for certain scenarios in military communications. The key observations in this study are as follows:

\begin{itemize}
	\item In satellite communications with small channel variations, it is not possible to support high overload factors compared to other scenarios, assuming the user received powers are close to each other. Also, superposing users with low code rate is not a valid solution, and NOMA schemes are required as the channel does not naturally separate the users. It is also observed that NOMA schemes are sensitive to CFO errors, and CFO correction is required in general. 
	\item For tactical area communications (BU), it is observed that NOMA schemes exploit the high frequency selectivity of the channel to create diversity and randomize the interference. The schemes are especially useful in high overload scenarios.
	\item For close-range/indoor communication scenarios (TDL-A), as the channel has larger coherence bandwidth compared to BU, less frequency diversity is observed. However, the channel creates a received power difference between the users so that they are separable in power domain. By using SIC based receivers, it is possible to support overload factors up to $500\%$ (and possibly higher) via NOMA schemes.
	\item S-OFDM provides good performance in low spectral efficiency and overload factors due to advanced MMSE-SIC-CRC receiver. However, in order to support larger overload values, specific NOMA schemes need to be used as illustrated in simulations. 
	\item SIMO technology is key in many scenarios as it provides an additional source of diversity.
	\item Advanced soft receivers can improve SCMA performance \cite{scma3}. For PDMA, it is known that MPA based receivers can provide performance gains over SIC-based receivers \cite{PDMA_TransComm}. For RDMA, the performance can be improved by using block based MMSE-SIC receivers at the expense of increased complexity.
	
\end{itemize}


%
%
%
%
%
%
%
%

\end{document}